\documentclass[%
    twocolumn,
    superscriptaddress,
    bitnotes,
    amsmath, amssymb, 
    aps, 
    prl,
    floatfix,
]{revtex4-2}

\usepackage{changes}
\usepackage{graphicx}       
\usepackage{dcolumn}        
\usepackage{bm}             
\usepackage{color}
\usepackage{subfigure}
\usepackage{appendix}
\usepackage[utf8]{inputenc}
\usepackage{hyperref}
\hypersetup{
    colorlinks=true,
    linkcolor=blue,
    filecolor=magenta,      
    urlcolor=cyan,
    pdfpagemode=FullScreen,
}

\definecolor{ROYALBLUE}{rgb}{0.25, 0.41, 0.88}
\definecolor{royalpurple}{rgb}{0.47, 0.32, 0.66}
\definecolor{coral}{rgb}{1.0, 0.5, 0.31}
\definecolor{chartreuse}{rgb}{0.55, 0.71, 0.0}

\definecolor{eggplant}{RGB}{180,33,147}

\begin{document}
\title{Robustness of Vacancy-Bound Non-Abelian Anyons in the Kitaev Model in a Magnetic Field}
\author{Bo Xiao}
\email{xiaob@ornl.gov}
\thanks{This manuscript has been authored by UT-Battelle, LLC, under contract DE-AC-05-00OR22725 with the US Department of Energy (DOE). The publisher acknowledges the US government license to provide public access under the DOE Public Access Plan (http://energy.gov/downloads/doe-public-access-plan).}
\affiliation{Materials Science and Technology Division, Oak Ridge National Laboratory, Oak Ridge, Tennessee 37831, USA}
\affiliation{Quantum Science Center, Oak Ridge, Tennessee 37831, USA}

\author{Gonzalo Alvarez}
\affiliation{Quantum Science Center, Oak Ridge, Tennessee 37831, USA}
\affiliation{Computational Science and Engineering Division, Oak Ridge National Laboratory, Oak Ridge, Tennessee 37831, USA}

\author{G\'{a}bor B. Hal\'{a}sz}
\affiliation{Materials Science and Technology Division, Oak Ridge National Laboratory, Oak Ridge, Tennessee 37831, USA}
\affiliation{Quantum Science Center, Oak Ridge, Tennessee 37831, USA}

\date{\today}
\begin{abstract}
    Non-Abelian anyons in quantum spin liquids (QSLs) provide a promising route to fault-tolerant topological quantum computation. In the exactly solvable Kitaev honeycomb model, such anyons of the QSL state can be bound to nonmagnetic spin vacancies and endowed with non-Abelian statistics by an infinitesimal magnetic field. Here, we investigate how this approach for stabilizing non-Abelian anyons extends to a finite magnetic field represented by a proper Zeeman term. Through large-scale density-matrix renormalization group (DMRG) simulations, we compute the vacancy-anyon binding energy as a function of magnetic field for both the ferromagnetic (FM) and antiferromagnetic (AFM) Kitaev models. We find that anyon binding remains robust within the entire QSL phase for the FM Kitaev model but breaks down already inside this phase for the AFM Kitaev model. To compute a binding energy several orders of magnitude below the magnetic energy scale, we introduce both a refined definition and an extrapolation scheme based on carefully tailored perturbations.
\end{abstract}

\maketitle
{\it Introduction.---}Quantum spin liquids (QSLs) are magnetic insulators characterized by long-range quantum entanglement~\cite{Balents2010,Savary2016,Zhou2017,Knolle2019,Broholm2020}. They support fractionalized quasiparticle excitations that interact via emergent gauge fields and possess nontrivial ``anyonic'' particle statistics beyond conventional bosons and fermions. For gapped non-Abelian QSLs, these anyonic quasiparticles---called non-Abelian anyons---enable topological quantum computation~\cite{Kitaev2003,Nayak2008}---an inherently fault-tolerant approach for storing and processing quantum information.

The celebrated Kitaev model on the honeycomb lattice~\cite{Kitaev2006} provides an exactly solvable realization of such a non-Abelian QSL. Indeed, while the ground state of the pure Kitaev model is a gapless $\mathbb{Z}_2$ QSL featuring gapless Majorana fermions coupled to gapped $\mathbb{Z}_2$ gauge fields, an external magnetic field immediately gaps out the Majorana fermions while turning the $\mathbb{Z}_2$ gauge fluxes into non-Abelian Ising anyons~\cite{Kitaev2006}. Moreover, the bond-dependent spin interactions of the Kitaev model naturally emerge~\cite{Jackeli2009,Chaloupka2010} in a class of spin-orbit-coupled honeycomb magnets commonly known as ``Kitaev materials''~\cite{Trebst2017,Hermanns2018,Takagi2019,Motome2019,Matsuda2025}. Therefore, these honeycomb magnets, including $\alpha$-RuCl$_3$~\cite{Plumb2014,Kubota2015,Sandilands2015,Sears2015,Majumder2015,Johnson2015,Sandilands2016,Banerjee2016,Leahy2017,Sears2017,Banerjee2017,Wolter2017,Baek2017,Do2017,Banerjee2018,Hentrich2018,Jansa2018,Kasahara2018,Widmann2019,Balz2019,Yamashita2020,Czajka2021,Yokoi2021,Bruin2022,Lefrancois2022,Czajka2023,Yang2023,Imamura2024a,Imamura2024b}, are promising candidates for realizing the non-Abelian QSL phase of the Kitaev model---along with its anyonic quasiparticles---in the presence of an applied magnetic field.

Recent years have seen numerous practical proposals for detecting, stabilizing, and controlling anyonic quasiparticles in this non-Abelian QSL phase~\cite{Aasen2020,Pereira2020,Konig2020,Udagawa2021,Klocke2021,Jang2021,Wei2021,Klocke2022,Liu2022,Bauer2023,Wei2023,Takahashi2023,Harada2023,Kao2024,Kao2024Dynamics,Halasz2024,Xiao2025}, culminating in concrete designs for prototype QSL-based topological qubits~\cite{Klocke2024}. Several of these proposals~\cite{Takahashi2023,Kao2024,Kao2024Dynamics,Halasz2024, Xiao2025} rely on the specific property of the pure Kitaev model that $\mathbb{Z}_2$ gauge fluxes are bound to both nonmagnetic spin vacancies~\cite{Willians2010,Willians2011,Kao2021} and Kondo impurities~\cite{Dhochak2010,Das2016,Vojta2016} that are equivalent to such spin vacancies in the limit of strong Kondo coupling. In an external magnetic field, each $\mathbb{Z}_2$ gauge flux then becomes a non-Abelian anyon, readily localized at the given spin vacancy. However, the stability of the resulting vacancy-anyon bound state has only been studied~\cite{Kao2024Dynamics,Halasz2024,Kao2021} in an exactly solvable extension of the Kitaev model where the magnetic field is represented by an effective three-spin interaction~\cite{Kitaev2006}. Although this exactly solvable model captures the universal properties of the non-Abelian QSL phase, it fails to account for some key nonuniversal features, e.g., the small size of the topological gap~\cite{Hickey2019,Zhang2022}, which may significantly impact the vacancy-anyon bound state.

\begin{figure*}
    \centering
    \includegraphics[width = 1\textwidth]{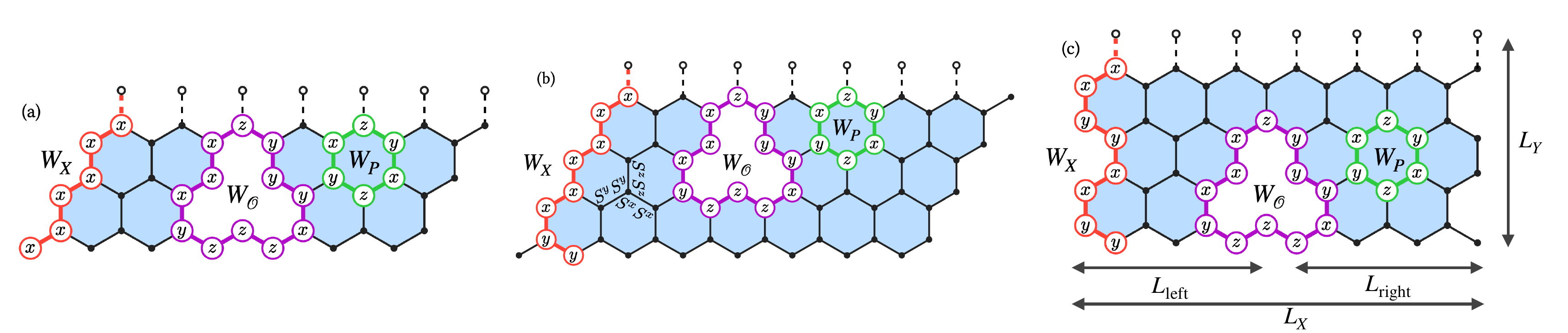}
    \caption{Honeycomb lattice in a narrow cylindrical geometry with (a) zigzag, (b) twisted, and (c) armchair boundary conditions, as well as a single spin vacancy in the center. 
    The length $L_X$, the width $L_Y$, and the partial lengths $L_{\rm left}$ and $L_{\rm right}$ of the cylinder are indicated, with (a) $L_Y=3$ and (b,c) $L_Y=4$. This figure shows $L_{\rm left}=L_{\rm right}=3$ and $L_X=L_{\rm left}+L_{\rm right}+1=7$ for compactness, but the calculations correspond to $L_{\rm left}=L_{\rm right}=7$ and $L_X=15$ in order to reduce edge effects. Each loop operator $W_{\mathcal{P}}$, $W_{\mathcal{O}}$, and $W_{X}$ introduced in the text is a simple product of appropriate spin operators ($x,y,z$) specified along the given green, purple, or red path. Anisotropic Kitaev spin interactions $S^{\alpha} S^{\alpha}$ along $\alpha = x,y,z$ bonds are also shown.}
    \label{fig:shcematic}
\end{figure*}

In this Letter, we numerically investigate the robustness of anyon binding at spin vacancies in the Kitaev model under an external magnetic field represented by a proper Zeeman term rather than the effective three-spin interaction. By employing the numerically exact density-matrix renormalization group (DMRG) method~\cite{White1992,White1993,Schollwock2011,Roman2019}, combined with an extrapolation scheme based on tailored perturbations, we calculate the vacancy-anyon binding energy as a function of the magnetic field inside the non-Abelian QSL phase of the Kitaev model. We observe a pronounced difference between the ferromagnetic (FM) and antiferromagnetic (AFM) Kitaev models, with robust anyon binding throughout the non-Abelian phase in the former case and a clear finite-field breakdown of the vacancy-anyon bound state in the latter case. For both signs of the Kitaev exchange, however, we find that anyon binding becomes weaker upon including a proper magnetic field, which is in stark contrast to the corresponding behavior for the effective three-spin interaction~\cite{Kao2024Dynamics}. Calculations across various system dimensions and boundary conditions suggest that our results extend all the way to the thermodynamic limit (TDL).

{\it Model and methods.---}We consider the Kitaev honeycomb model in an external magnetic field along the $[111]$ direction. The Hamiltonian of this model reads~\cite{Kitaev2006}
\begin{equation}
   H = -J \sum_{\langle i,j \rangle_{\alpha}} S_{i}^{\alpha} S_{j}^{\alpha}- h \sum_{i} \left(S_{i}^{x} + S_{i}^{y} + S_{i}^{z}\right), \label{eq:H}
\end{equation}
where $S_{i}^{\alpha}$ with $\alpha=x,y,z$ are conventional spin-$1/2$ operators at each lattice site $i$, and $\langle i,j \rangle_{\alpha}$ are $\alpha$ bonds of the lattice (see Fig.~\ref{fig:shcematic}) connecting pairs of sites $i$ and $j$. The first term represents the anisotropic Kitaev spin interactions, and the second term describes the Zeeman coupling of the spins to the magnetic field of magnitude $h$. At finite fields, the behavior of the model depends on the sign of the Kitaev exchange $J$~\cite{Zhu2018,Gohlke2018,Hickey2019}. For the FM Kitaev model ($J=+1$), the non-Abelian QSL directly transitions into a trivial polarized phase at a very small field. For the AFM Kitaev model ($J=-1$), however, the non-Abelian QSL persists up to a much larger field and is separated from the polarized phase by an intermediate QSL phase. Note that we set $|J|=1$ as the energy unit throughout this work.

DMRG simulations are performed on narrow cylinders of width $L_Y$ and length $L_X \gg L_Y$, using open (periodic) boundary conditions along the $X$ ($Y$) direction. To quantify finite-size effects originating from the small width, we implement distinct versions of the periodic boundary conditions for which loops winding around the short direction connect back to themselves through different twists along the long direction (see Fig.~\ref{fig:shcematic}). These boundary conditions interpolate between zigzag and armchair edges at the two ends of the cylinder and give access to different sets of parallel lines in momentum space~\cite{Gohlke2018}, thus giving rise to comparable but different finite-size effects. Our DMRG simulations use a maximum bond dimension $\chi_{\rm max} = 4500$ and a truncation error of $\mathcal{O}(10^{-10})$ to ensure numerical accuracy~\footnote{The perturbations used in our extrapolation scheme act like pinning fields, accelerating DRMG convergence. As a result, the bond dimension $\chi$ remains significantly below the maximum value $\chi_{\rm max}$ in most simulations, and $\chi_{\rm max} = 4500$ is sufficient to ensure good convergence.}.

{\it Spin vacancies and flux binding.---}In the pure Kitaev model at zero field ($h=0$), isolated spin vacancies are known to bind $\mathbb{Z}_{2}$ gauge fluxes of the QSL~\cite{Willians2010,Willians2011}. For an infinitesimal field ($h>0$), each $\mathbb{Z}_{2}$ gauge flux becomes a non-Abelian Ising anyon and remains bound to its parent spin vacancy. We aim to understand the stability of such a vacancy-flux bound state at finite fields by studying the energetics of flux binding at a single vacancy in the center of our cylinder (see Fig.~\ref{fig:shcematic}).

\begin{figure*}
    \centering
    \includegraphics[width=1.05\textwidth]{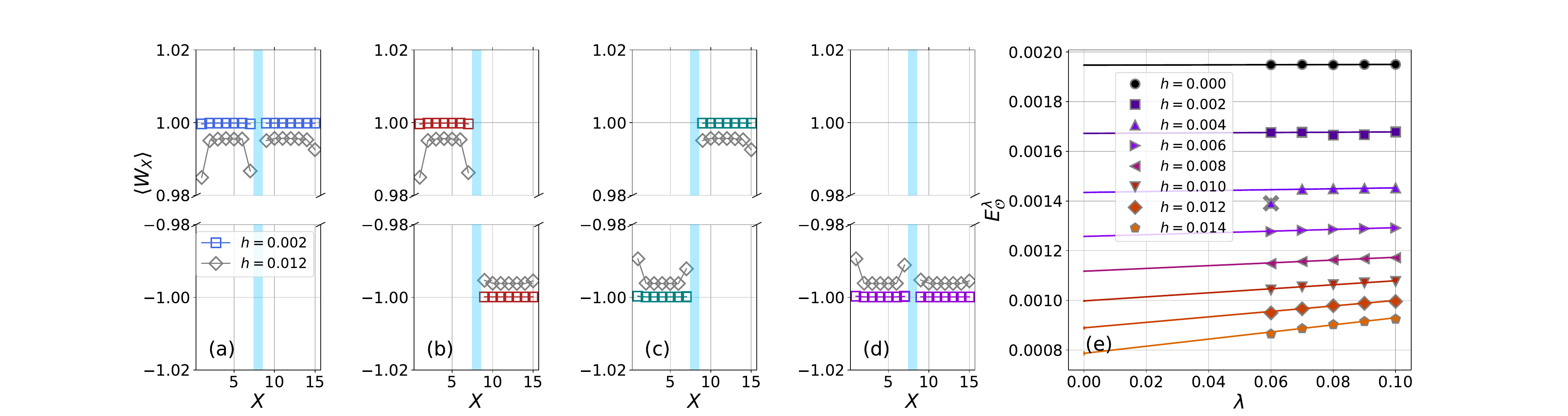}
    \caption{(a-d) Expectation value of the topological loop operator, $\langle W_X \rangle$, against the coordinate $1 \leq X \leq 15$ along the cylinder length for the perturbed Hamiltonians (a) $H^{\lambda}(+1, +1)$, (b) $H^{\lambda}(+1, -1)$, (c) $H^{\lambda}(-1, +1)$, and (d) $H^{\lambda}(-1, -1)$ in Eq.~(\ref{eq:H'}) with perturbation strength $\lambda=0.06$, FM Kitaev exchange $J=1$, as well as magnetic fields $h=0.002$ (colorful solid symbols) and $h=0.012$ (gray empty symbols) on a narrow cylinder with $L_X=15$, $L_Y=3$, $L_{\rm left}=L_{\rm right}=7$, and twisted boundary conditions. The light blue shading marks the non-existent topological loop operator going through the vacancy at coordinate $X=X_{\mathcal{O}}=8$. (e) Extraction of the unperturbed energy coefficient $E_{\mathcal{O}}$ in Eq.~(\ref{eq:binding-2}) through a linear extrapolation of the perturbed energy coefficient $E_{\mathcal{O}}^{\lambda}$ in Eq.~(\ref{eq:binding-4}) from $0.06 \leq \lambda \leq 0.1$ to $\lambda = 0$ for a variety of magnetic fields $0 \leq h \leq 0.014$. The gray cross marks an obvious outlier value that is excluded from the extrapolation.
    }
    \label{fig:loop}
\end{figure*}

We start by providing an appropriate definition of the vacancy-flux binding energy in our narrow cylindrical geometry. At zero field ($h=0$), the $\mathbb{Z}_{2}$ gauge fluxes correspond to conserved loop operators $W$ that are simple products of spin operators along closed loops of the lattice (see Fig.~\ref{fig:shcematic}) and take eigenvalues $\overline{W}=\pm1$~\cite{Kitaev2006}. These loop operators include the regular plaquette operators $W_{\mathcal{P}}$, the vacancy-plaquette operator $W_{\mathcal{O}}$, and the topological loop operators $W_X$, labeled by their coordinate $X$ along the cylinder length. Assuming no flux excitations on regular plaquettes (i.e., $\overline{W}_{\mathcal{P}}=+1$), there are only two independent loop eigenvalues, $\overline{W}_{\mathrm{left}}\equiv\overline{W}_{X<X_{\mathcal{O}}}$ and $\overline{W}_{\mathrm{right}}\equiv\overline{W}_{X>X_{\mathcal{O}}}$, that correspond to topological loops on the left and right sides of the vacancy at $X=X_{\mathcal{O}}$, respectively, with the vacancy-plaquette eigenvalue given by their product: $\overline{W}_{\mathcal{O}} = \overline{W}_{\mathrm{left}} \overline{W}_{\mathrm{right}}$. The four topological flux sectors labeled by $\overline{W}_{\mathrm{left}}=\pm1$ and $\overline{W}_{\mathrm{right}}=\pm1$ each have a lowest-energy state (``ground state''), and the four individual ground-state energies can be written as
\begin{align}
    E_{\overline{W}_{\mathrm{left}}, \overline{W}_{\mathrm{right}}} = & \,\, E_{\rm const} + E_{\rm left} \overline{W}_{\rm left} + E_{\rm right} \overline{W}_{\rm right} \nonumber \\
    &+ E_{\mathcal{O}} \overline{W}_{\rm left} \overline{W}_{\rm right}. \label{eq:E}
\end{align}
The coefficients $E_{\rm const}$, $E_{\rm left}$, $E_{\rm right}$, and $E_{\mathcal{O}}$ can be obtained by computing the ground-state energy for each topological sector and solving the resulting four equations for the four coefficients. The scalings of these coefficients with the system dimensions (see End Matter) reveal that $E_{\rm left}$ and $E_{\rm right}$ are emphasized in our narrow cylindrical geometry as---in contrast to the TDL---the topological loop operators $W_X$ are actually local (see Fig.~\ref{fig:shcematic}).

While $E_{\rm left}$ and $E_{\rm right}$ depend linearly on $L_X$ and decay to zero with $L_Y$ (see End Matter), $E_{\mathcal{O}}$ has a much weaker dependence on the cylinder dimensions and directly corresponds to half the energy difference between $\overline{W}_{\mathcal{O}} = +1$ (no vacancy-flux binding) and $\overline{W}_{\mathcal{O}} = -1$ (vacancy-flux binding) even in the TDL. Hence, the most natural way to infer the binding energy in the TDL is to obtain this coefficient---explicitly given by 
\begin{align}
\begin{split}
   E_{\mathcal{O}} &= \frac{1}{4} \Big[E_{+1, +1} + E_{-1, -1} - E_{+1, -1} - E_{-1, +1} \Big] \label{eq:binding-2}
\end{split}
\end{align}
in terms of $E_{\overline{W}_{\mathrm{left}}, \overline{W}_{\mathrm{right}}}$ [see Eq.~(\ref{eq:E})]---for each finite geometry and then extrapolate to the TDL:
\begin{equation}
   E_{\rm binding} = -2 \lim_{L_Y \to \infty} \lim_{L_X \to \infty} E_{\mathcal{O}}.
   \label{eq:binding-1}
\end{equation}
Here it is implicitly assumed that the vacancy is near the center of the cylinder so that the partial lengths $L_{\rm left}$ and $L_{\rm right}$ (see Fig.~\ref{fig:shcematic}) both diverge for $L_X \to \infty$.

We stress that our refined definition of the vacancy-flux binding energy in Eqs.~(\ref{eq:binding-1}) and (\ref{eq:binding-2}) is critical for making a connection from our narrow cylindrical geometry to the TDL. More naive definitions (see End Matter) would not extract $E_{\mathcal{O}}$ on its own but together with $E_{\rm left}$ or $E_{\rm right}$ that are artifacts of the cylindrical geometry and may overshadow $E_{\mathcal{O}}$. We further note that, using our refined definition, we find $E_{\rm binding} = -0.006762 \pm 0.000009$ for the binding energy in the TDL (see End Matter)---which is consistent with Ref.~\onlinecite{Willians2010} but has an order of magnitude better accuracy.

{\it Finite-field binding energy.---}At finite fields ($h>0$), the original versions of the topological loop operators, $W_X$, no longer commute with the Hamiltonian $H$. Nevertheless, inside the non-Abelian QSL phase, the underlying topological order implies modified versions of these topological loop operators, $W_X'(h)$, that can be obtained perturbatively in the field $h$ and remain conserved quantities even at $h>0$~\cite{Savary2016}. Given that the four combinations of the associated eigenvalues $\overline{W}_{\mathrm{left}}\equiv\overline{W}_{X<X_{\mathcal{O}}}'(h)=\pm1$ and $\overline{W}_{\mathrm{right}}\equiv\overline{W}_{X>X_{\mathcal{O}}}'(h)=\pm1$~\footnote{Note that, for simplicity, we use the same notation for the topological eigenvalues $\overline{W}_{\mathrm{left/right}}$ as before but implicitly understand that they are eigenvalues of $W_X$ at $h=0$ and eigenvalues of $W_X'(h)$ at $h>0$.} still correspond to four disconnected topological sectors with well-defined ground-state energies $E_{\overline{W}_{\mathrm{left}}, \overline{W}_{\mathrm{right}}}$, we can then continue to define the vacancy-flux binding energy through Eqs.~(\ref{eq:binding-1}) and (\ref{eq:binding-2}).

\begin{figure*}
    \centering 
    \includegraphics[width=1\textwidth]{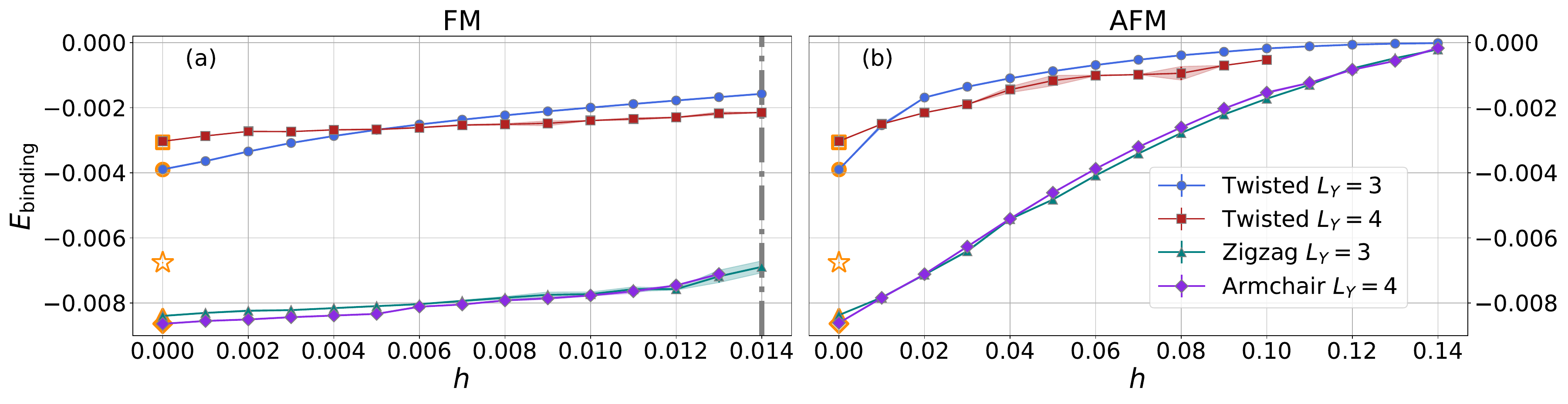} 
    \caption{Vacancy-flux binding energy, $E_{\rm binding} = -2 E_{\mathcal{O}}$, as a function of the external magnetic field $h$ for the (a) FM and (b) AFM Kitaev model on a width $L_Y=3$ cylinder with zigzag (teal triangles) and twisted (blue circles) boundary conditions as well as a width $L_Y=4$ cylinder with armchair (purple diamonds) and twisted (red squares) boundary conditions. The orange triangle, circle, diamond, and square represent the corresponding exact values at $h=0$, while the orange star shows the exact value at $h=0$ in the TDL. For the FM Kitaev model, the dash-dotted line marks the approximate critical field, $h_C^{\rm FM} \approx 0.014$, for the transition out of the non-Abelian QSL phase~\cite{Gohlke2018}. For the AFM Kitaev model, the analogous critical field, $h_C^{\rm AFM} \approx 0.22$, lies outside the simulated field range. The cylinder dimensions are $L_X=15$ and $L_{\rm left}=L_{\rm right}=7$ throughout this figure.}
    \label{fig:binding_energy}
\end{figure*}

To extract the four individual ground-state energies $E_{\overline{W}_{\mathrm{left}}, \overline{W}_{\mathrm{right}}}$, we employ an extrapolation scheme where each of the ground states is turned into the overall ground state by applying an appropriate perturbation that approximately commutes with the Hamiltonian and energetically favors the given ground state. For a sufficiently small field $h>0$ inside the non-Abelian QSL phase, the most natural perturbations are the topological loop operators $W_X$ at $h=0$ whose expectation values are still approximately $\pm 1$ for $h>0$. The appropriate perturbed Hamiltonian for extracting the ground state of a given topological sector $(\overline{W}_{\mathrm{left}}, \overline{W}_{\mathrm{right}})$ is then
\begin{equation}
   H^{\lambda} (\overline{W}_{\mathrm{left}}, \overline{W}_{\mathrm{right}}) = H - \lambda H_X (\overline{W}_{\mathrm{left}}, \overline{W}_{\mathrm{right}}) - \lambda \sum_{\mathcal{P}} W_{\mathcal{P}}, \label{eq:H'}
\end{equation}
where the first term $H$ is the unperturbed Hamiltonian in Eq.~(\ref{eq:H}), the second term with
\begin{equation}
   H_X (\overline{W}_{\mathrm{left}}, \overline{W}_{\mathrm{right}}) = \overline{W}_{\rm left} \sum_{X < X_{\mathcal{O}}} W_X + \overline{W}_{\rm right} \sum_{X > X_{\mathcal{O}}} W_X \label{eq:HX}
\end{equation}
makes the specific topological sector energetically favorable, while the third term suppresses spurious flux excitations on regular plaquettes $\mathcal{P}$ induced by the small cylinder width. Above a finite critical perturbation strength $\lambda$, the overall ground state of each perturbed Hamiltonian $H^{\lambda} (\overline{W}_{\mathrm{left}}, \overline{W}_{\mathrm{right}})$ is in the desired topological sector, $(\overline{W}_{\mathrm{left}}, \overline{W}_{\mathrm{right}})$, and the perturbed ground-state energy of this topological sector, $E^{\lambda}_{\overline{W}_{\mathrm{left}}, \overline{W}_{\mathrm{right}}}$, can thus be directly calculated by DMRG as the overall ground-state energy. Since $E^{\lambda}_{\overline{W}_{\mathrm{left}}, \overline{W}_{\mathrm{right}}}$ varies smoothly with $\lambda$, even below the critical value where it may no longer be the overall ground-state energy, one can then extrapolate to $\lambda=0$ from a finite range above the critical value in order to obtain the corresponding ground-state energy $E_{\overline{W}_{\mathrm{left}}, \overline{W}_{\mathrm{right}}}$ for the unperturbed model.

Figure \ref{fig:loop} illustrates this extrapolation scheme through a concrete example. As depicted in Figs.~\ref{fig:loop}(a-d), the expectation values of the topological loop operators $W_X$ in the overall ground state, as obtained by DMRG, already have the desired sign structure in each topological sector for perturbation strength $\lambda = 0.06$. Hence, we can extract the coefficient $E_{\mathcal{O}}$ in Eq.~(\ref{eq:binding-2}) by directly computing the perturbed ground-state energy $E^{\lambda}_{\overline{W}_{\mathrm{left}}, \overline{W}_{\mathrm{right}}}$ of each topological sector $(\overline{W}_{\mathrm{left}}, \overline{W}_{\mathrm{right}})$ as the overall ground-state energy of the perturbed Hamiltonian $H^{\lambda} (\overline{W}_{\mathrm{left}}, \overline{W}_{\mathrm{right}})$ for a range of perturbation strengths, $0.06 \leq \lambda \leq 0.1$, and then extrapolating 
\begin{align}
\begin{split}
   E_{\mathcal{O}}^{\lambda} &= \frac{1}{4} \Big[E^{\lambda}_{+1, +1} + E^{\lambda}_{-1, -1} - E^{\lambda}_{+1, -1} - E^{\lambda}_{-1, +1} \Big]
\end{split} \label{eq:binding-4}
\end{align}
to $\lambda=0$ in order to find $E_{\mathcal{O}}^{\phantom{\lambda}} \equiv E_{\mathcal{O}}^{\lambda=0}$. Figure \ref{fig:loop}(e) shows that a simple linear extrapolation works well in the entire range $0.06 \leq \lambda \leq 0.1$ and can thus be reliably used to determine $E_{\mathcal{O}}$. The suitability of the linear extrapolation stems from the fact that the topological loop operators $W_X$ at $h=0$ have approximately constant expectation values $\pm 1$ (i.e., approximately commute with the Hamiltonian) even for the largest fields in the non-Abelian QSL phase [see Figs.~\ref{fig:loop}(a-d)]. Occasional ``outliers'' in the $E_{\mathcal{O}}^{\lambda}$ values of Fig.~\ref{fig:loop}(e) that appear due to incomplete DMRG convergence can be readily identified and excluded, as described in the Supplemental Material (SM)~\footnote{See Supplemental Material at [URL] for a detailed description of the extrapolation procedure used to extract the binding energy in both the ferromagnetic and antiferromagnetic Kitaev models at representative values of the magnetic field. The Supplement further details the systematic protocol employed to identify and discard outliers in the extrapolation process.}. The complete set of results for $E_{\mathcal{O}}^{\lambda}$ at representative values of the perturbation strength, including the outliers, are also provided in the SM.

{\it Results and discussion.---}Figure \ref{fig:binding_energy} shows our results for the vacancy-flux binding energy, $E_{\rm binding} = -2 E_{\mathcal{O}}$, against the magnetic field $h$ for both FM and AFM Kitaev models with a single vacancy in the center of the cylinder. For each sign of the Kitaev exchange, the four curves correspond to different cylinder widths $L_Y$ and/or boundary conditions. At zero field, the binding energy is identical for the two signs, and can be readily obtained through the exact solution of the Kitaev model~\cite{Kitaev2006}. This benchmark is quantitatively reproduced by our DMRG results for both signs across all finite geometries, demonstrating the accuracy and reliability of the method.

As shown in Fig.~\ref{fig:binding_energy}, the binding energy at zero field is consistently negative across all cylinder widths and boundary conditions, confirming that flux binding to the vacancy is energetically favorable. Despite the significant scattering among the binding energies found for the various geometries, all values have the same order of magnitude and are between $-0.009$ and $-0.003$. Reassuringly, the binding energy in the TDL, $E_{\rm binding} \approx -0.00676$~\cite{Willians2010}, is near the center of this range, indicating that a rough ``average'' of our four finite-size binding energies is a good representation of the TDL value. Therefore, any common trend observed in the binding energy for all four finite geometries is expected to apply even in the TDL.

At finite fields, the behavior of the binding energy exhibits striking differences between the FM and AFM Kitaev models. For the FM Kitaev model, the binding energy decreases in magnitude but remains clearly negative all the way to the phase transition out of the non-Abelian QSL at $h_C^{\rm FM} \approx 0.014$~\cite{Gohlke2018}, indicating that the flux remains robustly bound to the vacancy in the presence of a finite field. For the AFM Kitaev model, in contrast, the binding energy drops more rapidly and already approaches zero at $h \approx 0.14$, i.e., significantly before the phase transition out of the non-Abelian QSL at $h_C^{\rm AFM} \approx 0.22$~\cite{Gohlke2018}. In other words, flux binding is energetically favorable in the weak-field regime but is destabilized as the field strengthens. Since the same trends for the two signs of the Kitaev exchange are observed for all four finite geometries, we expect that the marked distinction between the FM and AFM Kitaev models persists even in the TDL.

{\it Summary.---}We numerically computed the binding energy between a non-Abelian Ising anyon and a nonmagnetic spin vacancy in both the FM and AFM Kitaev models under an external magnetic field. We observed a striking difference between the two cases, with robust anyon binding within the entire non-Abelian QSL phase for the FM Kitaev model but only up to a finite field inside this phase for the AFM Kitaev model. Our results highlight the robustness of vacancy-bound non-Abelian anyons in realistic models, thus providing a promising foundation for qubit architectures based on QSL materials.


\vskip0.05in
{\it Acknowledgments---}We are grateful to Lukas Weber for assistance in creating the schematic figure. We also thank Satoshi Fujimoto, Wen-Han Kao, Natalia Perkins, Miles Stoudenmire, and Masahiro Takahashi for valuable discussions. This material is based upon work supported by the U.S. Department of Energy, Office of Science, National Quantum Information Science Research Centers, Quantum Science Center. This research used resources of the National Energy Research Scientific Computing Center (NERSC), a DOE Office of Science User Facility supported by the Office of Science of the U.S. Department of Energy under Contract No. DE-AC02-05CH11231 using NERSC award ASCR-ERCAP0032461.

\vskip0.05in
{\it Data availability---}The data that support the findings of this Letter are not publicly available. The data are available from the authors upon reasonable request.

\bibliography{kitaev}

\clearpage
\appendix
\section{End Matter} \label{appendix:end_matter}

\subsection{\label{sec:refined_binding} System-size dependence of the energy coefficients}

\begin{figure*}
    \centering
    \includegraphics[width=1\textwidth]{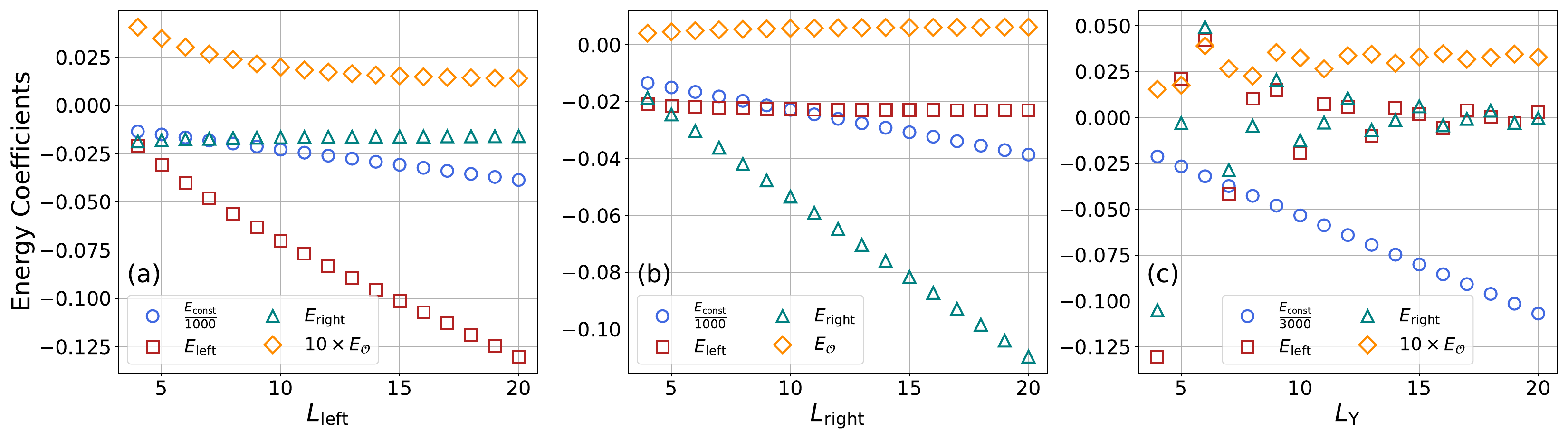}
    \caption{Energy coefficients $E_{\rm const}$, $E_{\rm left}$, $E_{\rm right}$, and $E_{\mathcal{O}}$ [see Eq.~(\ref{eq:E})] against (a) the partial cylinder length on the left side of the vacancy, $L_{\rm left}$, for fixed $L_{\rm right} = L_Y = 4$, (b) the partial cylinder length on the right side of the vacancy, $L_{\rm right}$, for fixed $L_{\rm left} = L_Y = 4$, and (c) the cylinder width $L_Y$ for fixed $L_{\rm left} = L_{\rm right} = 20$. These coefficients are calculated through the exact solution of the pure Kitaev model at zero magnetic field ($h=0$). Note that certain coefficients are plotted with an appropriate scaling factor (see the legend of each panel) in order to be comparable with other coefficients.
    }
    \label{fig:E}
\end{figure*}

Figure \ref{fig:E} illustrates how the coefficients $E_{\rm const}$, $E_{\rm left}$, $E_{\rm right}$, and $E_{\mathcal{O}}$ in Eq.~(\ref{eq:E}) depend on the cylinder width $L_Y$ as well as the partial cylinder lengths $L_{\rm left}$ and $L_{\rm right}$ (see Fig.~\ref{fig:shcematic}). Since the coefficient $E_{\rm const}$ determines the overall ground-state energy, it scales linearly with $L_Y$, $L_{\rm left}$, and $L_{\rm right}$ alike. In contrast, the coefficient $E_{\rm left}$ ($E_{\rm right}$) scales linearly with $L_{\rm left}$ ($L_{\rm right}$) but is largely independent of $L_{\rm right}$ ($L_{\rm left}$). The reason is that the number of independent topological loop operators with the same eigenvalue $\overline{W}_{\mathrm{left}}$ ($\overline{W}_{\mathrm{right}}$) on the left (right) side of the vacancy is precisely the appropriate partial length $L_{\rm left}$ ($L_{\rm right}$). At the same time, both $E_{\rm left}$ and $E_{\rm right}$ decay to zero as $L_Y$ is increased, which indicates that these coefficients are emphasized in our narrow cylindrical geometry. Most importantly, the final coefficient $E_{\mathcal{O}}$ does not scale linearly with any system dimension and appears to converge to a finite value in the TDL.

\subsection{\label{sec:refined_binding} Refined definition of the binding energy}

The refined definition of the vacancy-flux binding energy in Eqs.~(\ref{eq:binding-1}) and (\ref{eq:binding-2}) is critical for establishing a connection between our narrow cylindrical geometry and the TDL. Using this definition, the binding energy is obtained as a linear combination of four distinct ground-state energies. In contrast, naive definitions adopted directly from the TDL would take a simple difference between two ground-state energies:
\begin{align}
\begin{split}
    E_{\rm binding}' &= \lim_{L_Y \to \infty} \lim_{L_X \to \infty} \left[ E_{-1,+1} - E_{+1,+1} \right] \\
    &= -2 \lim_{L_Y \to \infty} \lim_{L_X \to \infty} \left[ E_{\mathcal{O}} + E_{\rm left} \right], \label{eq:binding-3} \\
    E_{\rm binding}'' &= \lim_{L_Y \to \infty} \lim_{L_X \to \infty} \left[ E_{+1,-1} - E_{+1,+1} \right]  \\
    &= -2 \lim_{L_Y \to \infty} \lim_{L_X \to \infty} \left[ E_{\mathcal{O}} + E_{\rm right} \right]. 
\end{split}
\end{align}
Since $E_{\rm left}$ ($E_{\rm right}$) scales linearly with $L_{\rm left}$ ($L_{\rm right}$) for a narrow cylinder, these naive definitions give binding energies that depend heavily on the precise cylinder length and are typically an order of magnitude larger than the binding energy in the TDL.

\begin{figure}
     \hspace{-0.2in}
     \includegraphics[width=0.5\textwidth]{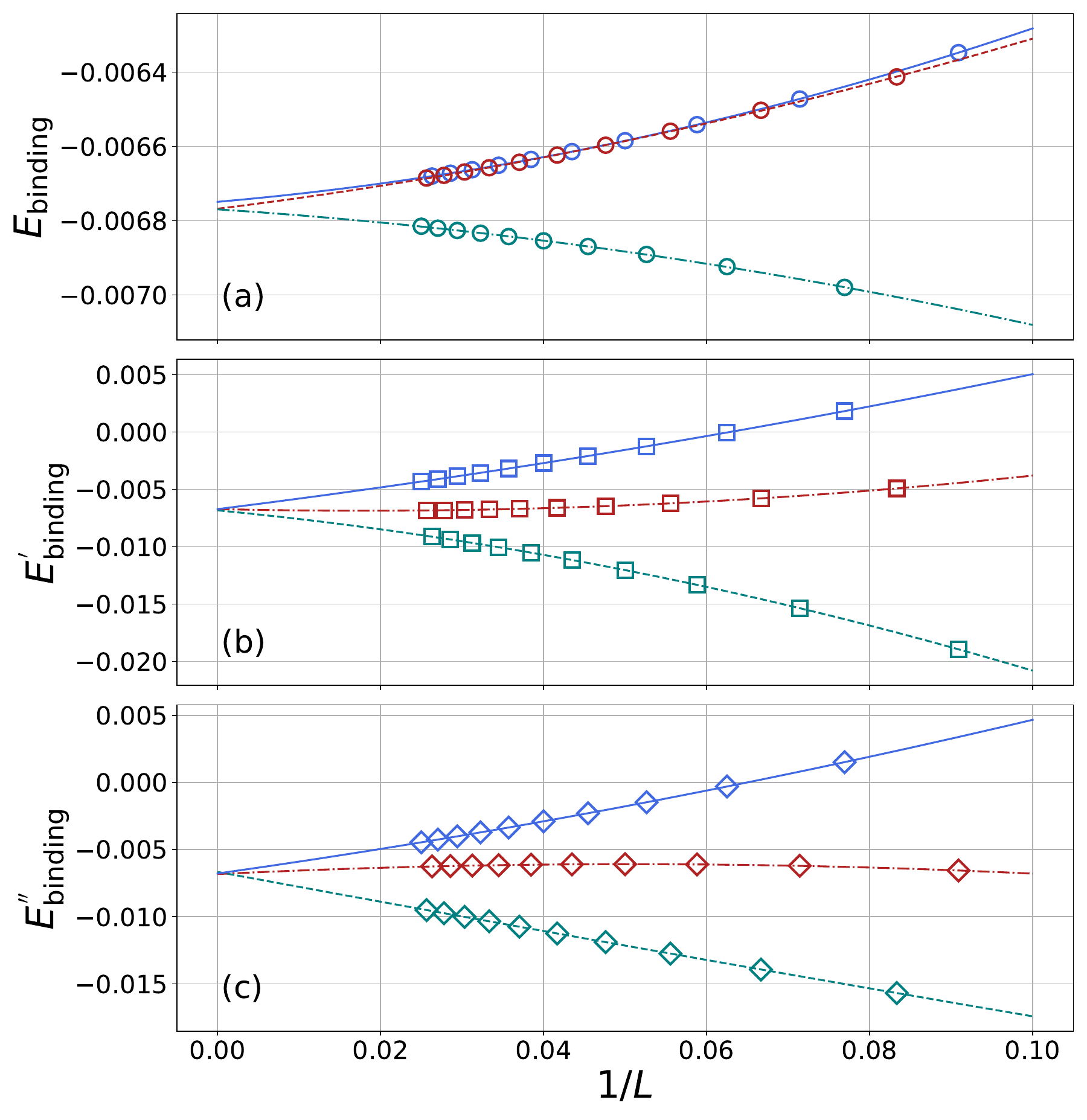}
     \caption{Extraction of the vacancy-flux binding energy in the TDL of the pure Kitaev model via a standard finite-size scaling in the cylinder dimension $L = L_{\rm left} = L_{\rm right} = L_Y$ with $11 \leq L \leq 40$, using (a) the refined definition in Eqs.~(\ref{eq:binding-1}) and (\ref{eq:binding-2}) as well as (b,c) the naive definitions in Eq.~(\ref{eq:binding-3}). The solid, dashed, and dotted lines represent quadratic extrapolations in $1/L$ from finite-size binding energies with specific moduli of $L$ with respect to $3$, as indicated by different colors.}
     \label{fig:binding_exact}
\end{figure}

To demonstrate the utility of our refined definition in Eqs.~(\ref{eq:binding-1}) and (\ref{eq:binding-2}), even away from the narrow cylindrical regime, we employ this refined definition to determine the vacancy-flux binding energy in the TDL up to one more significant figure than in Ref.~\onlinecite{Willians2010}. By setting the cylinder dimensions to be $L_{\rm left} = L_{\rm right} = L_Y = L$, and performing a standard finite-size scaling, as shown in Fig.~\ref{fig:binding_exact}(a), we extract the binding energy in the TDL to be $E_{\rm binding} = -0.006762 \pm 0.000009$. This value is consistent with the one previously found in Ref.~\onlinecite{Willians2010} but has an order of magnitude better accuracy~\footnote{Ref.~[63] uses a different convention for the spin Hamiltonian, and our value for the binding energy corresponds to $E_{\rm binding} = -0.02705 \pm 0.00004$ in their convention.}. If we instead use the naive definitions in Eq.~(\ref{eq:binding-3}), analogous finite-size scalings in Figs.~\ref{fig:binding_exact}(b,c) yield $E_{\rm binding}' = -0.00676 \pm 0.00005$ as well as $E_{\rm binding}'' = -0.00676 \pm 0.00007$, which are comparable to Ref.~\onlinecite{Willians2010} in terms of their accuracy.

\begin{figure*}
     \centering
     \includegraphics[width=1\textwidth]{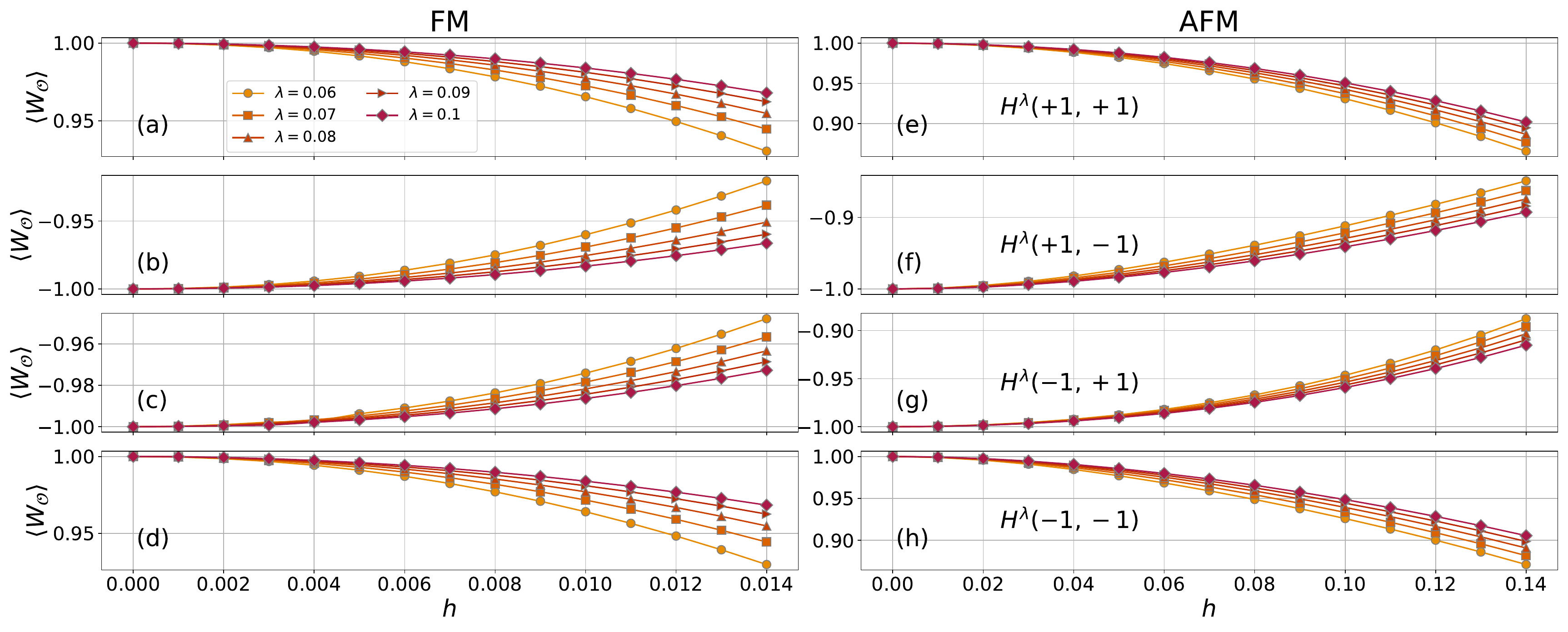}
     \caption{Expectation value of the vacancy-plaquette operator, $\langle W_{\mathcal{O}} \rangle$, against the magnetic field $h$ for the perturbed Hamiltonians (a,e) $H^{\lambda} (+1, +1)$, (b,f) $H^{\lambda} (+1, -1)$, (c,g) $H^{\lambda} (-1, +1)$, and (d,h) $H^{\lambda} (-1, -1)$ in Eq.~(\ref{eq:H'}) with a range of perturbation strengths $0.06 \leq \lambda \leq 0.1$ (different colors) as well as (a-d) FM Kitaev exchange $J = 1$ and (e-h) AFM Kitaev exchange $J = -1$ on a narrow cylinder with $L_X = 15$, $L_Y = 3$, $L_{\rm left} = L_{\rm right} = 7$, and twisted boundary conditions.}
     \label{fig:order_parameter}
\end{figure*}

\subsection{\label{sec:order_parameter} Expectation value of the vacancy-plaquette operator}

To elucidate the nature of flux binding at a finite field $h$, we also consider the expectation value of the vacancy-plaquette operator, $\langle W_{\mathcal{O}} \rangle$, for the four distinct topological sectors. At zero field ($h=0$), the vacancy-plaquette operator $W_{\mathcal{O}}$ is a conserved quantity, and its expectation value is simply $\langle W_{\mathcal{O}} \rangle = \overline{W}_{\mathcal{O}} = \overline{W}_{\mathrm{left}} \overline{W}_{\mathrm{right}} = \pm 1$ for each topological sector, with the eigenvalue $+1$ ($-1$) indicating the absence (presence) of flux binding. At finite fields ($h>0$), however, the conserved eigenvalue $\overline{W}_{\mathcal{O}} = \pm 1$ corresponds to a modified version of this plaquette operator, as described for the analogous topological operators in the main text, and the expectation value of the original plaquette operator $W_{\mathcal{O}}$ thus deviates from $\pm 1$. Nevertheless, as plotted in Fig.~\ref{fig:order_parameter}, the expectation value $\langle W_{\mathcal{O}} \rangle$ remains close to $\pm 1$ in each topological sector even for the largest fields considered, which means that a positive (negative) expectation value $\langle W_{\mathcal{O}} \rangle$ still indicates the absence (presence) of flux binding. We remark that, while the curves in Fig.~\ref{fig:order_parameter} correspond to finite values of the perturbation strength $\lambda$ [see Eq.~(\ref{eq:H'})], a rough extrapolation reveals that the same conclusion also holds for the unperturbed model at $\lambda=0$.

\end{document}